\documentclass[aps,prb,twocolumn,showpacs,amsmath,amssymb,unsortedaddress,floatfix]{revtex4-1}

\usepackage{graphicx}
\usepackage{tabularx}
\usepackage{color}
\usepackage{natbib}

\newcommand{\Ang}{\,\mathrm{\AA}}
\newcommand{\ket}[1]{\bigl| {#1} \bigr\rangle}

\def\eV{\,\textrm{eV}}

\def\nm{\,\textrm{nm}}
\def\etal{\textit{et al.}{}}

\newcommand{\hbn}{{\it h}-BN}
\newcommand{\eBN}{\ensuremath{ \epsilon_{\mathrm{BN}}}}
\newcommand{\grb}{$\mathrm{Gr_B}$}
\newcommand{\grt}{$\mathrm{Gr_T}$}

\newcommand{\Vg} {\ensuremath{{V_g}}}
\newcommand{\Dg} {\ensuremath{{D_g}}}
\newcommand{\Eint} {\ensuremath{{E_{\mathrm{inside}}}}}
\newcommand{\rhoif} {\ensuremath{ \Delta\rho_\mathrm{intf}}}
\newcommand{\rhob} {\ensuremath{ \Delta\rho_\mathrm{bulk}}}
\newcommand{\braket}[2]{\ensuremath{\left\langle#1 \vphantom{#2}\middle|  #2 \vphantom{#1}\right\rangle}}

\begin{document}

\title{ First-principles Simulations of a Graphene Based Field-Effect Transistor}

\author{Yun-Peng Wang and Hai-Ping Cheng } 
\email[Corr. author: Hai-Ping Cheng,  ]{cheng@qtp.ufl.edu}
\affiliation{Department of Physics, University of Florida,  Gainesville, Florida 32611, USA}

\begin{abstract}

We improvise a novel approach to carry out first-principles simulations 
of graphene-based vertical field effect tunneling transistors that consist of
a graphene$|$\hbn$|$graphene multilayer structure.
Within the density functional theory framework, we exploit the effective screening medium (ESM) method
to properly treat boundary conditions for electrostatic potentials
and investigate the effect of gate voltage.
The distribution of free carriers and the band structure of both top and bottom graphene layers are calculated self-consistently.
The dielectric properties of \hbn{} thin films sandwiched between graphene layers
are computed layer-by-layer following the theory of microscopic permittivity.
We find that the permittivities of BN layers are very close to that of crystalline \hbn.
The effect of interface with graphene on the dielectric properties of \hbn{} is weak,
according to an analysis on the interface charge redistribution.
\end{abstract}

\maketitle

\section{introduction}

\begin{figure}[b]
\begin{center}
\includegraphics[width=0.7 \linewidth]{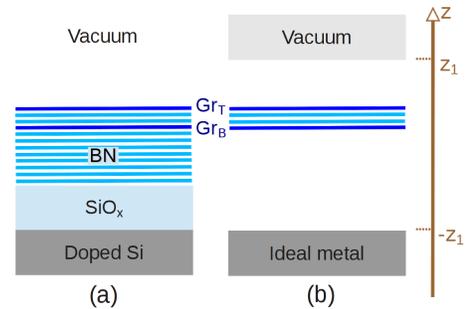}
\caption{
\label{fig:struct}
(Color Online)
(a) Schematic of a graphene-based field effect tunneling transistor.
The core graphene$|$\hbn$|$graphene structure is separated from the doped Si (which serves as the gate electrode)
by \hbn{} crystal and $\mathrm{SiO}_x$ slabs (which serve as dielectrics).
The graphene layers closer to and further away from the gate electrode are labelled as \grb{} and \grt, respectively.
(b) The model used in our simulations.
The dielectrics (\hbn\ and $\mathrm{SiO}_x$) are replaced by vacuum. 
The core graphene$|$\hbn$|$graphene structure is embedded betwen a semi-infinite vacuum ($z>z_1$)
and an ideally metallic medium ($z<-z_1$, which replaces the doped Si gate).
}
\end{center}
\end{figure}

Graphene-based field-effect transistors have been synthesized,\cite{Science-graphene} 
but the ratio of high and low resistances (switching ratio) is limited (less than a factor of $10$) 
because of the zero-gap band structure of pristine graphene.
A possible solution is to introduce an energy gap into graphene by using, for instance,
bilayer graphene,\cite{PhysRevLett.99.216802,Nat.Mater.7.151}
nanoribbons,\cite{PhysRevLett.98.206805} or chemical derivatives.\cite{Science.323.610}
Recently, Britnell {\it et al.}\cite{Science.335.947,NNANO.8.100,NCOMMS.4.1794}
reported an alternative transistor architecture, a vertical field effect tunneling transistor (FETT), 
exhibiting a fairly high ($\sim 50$) switching ratio.
In FETT devices, a graphene$|$barrier$|$graphene trilayer is utilized as the current-carrying channel, 
as in Fig.~\ref{fig:struct}(a); 
the electric current flows into the channel through one graphene layer and out through the other.
Inside the trilayer, electrons cross the thin barrier via tunneling.
The chemical potentials of the two graphene layers and the tunneling conductance can be tuned by gate voltages.
The bottom graphene layer (closer to gate electrode, \grb)
can only partially screen the gate voltage because of
the low density of states (DOS) near the Dirac point,
and so the top graphene layer (away from gate electrode, \grt) is also affected by the gate voltage.

In a typical field-effect device, the current-carrying channel 
is separated by a thick dielectric from a metallic gate electrode.
A gate voltage applied between the gate electrode and the current-carrying channel
controls the free carrier density in the channel.
The spatial distribution of free carriers can be obtained by solving the electrostatic Poisson equation.
Macroscopic physical quantities of device components are used as parameters for Poisson's equation,
including dielectric constants and electron affinities of dielectrics and work functions of gate electrodes, {\it etc}.
\cite{APL.90.143108}
%However, as the size of a device shrinks down to nanometer scale, macroscopic electrostatics faces a difficult challenge,
%due to ill-defined macroscopic physical quantities in heterostructured nanoscale electronic devices;%
%\cite{Nature.402.273,Nature.407.57,Nature.417.722}
%for instance, dielectric properties of an interface departure significantly from the corresponding bulk.\cite{PhysRevLett.91.267601}
%%{\color{blue}
However, the extrapolation of the macroscopic theory of electrostatics to the nanometer scale is unjustified,
because interfaces often exhibit dielectric screening properties different from the bulk.\cite{Jackson}
%As the ratio of interface to volume increases,
%the dielectric screening of a system can no longer be accurately described using the bulk dielectric constant.
%The applicability of the macroscopic theory depends on the amount of deviation of interfaces from the bulk
%and on the ratio of interface to volume.
%The macroscopic theory fails when the size of a system is not sufficiently larger than the thickness of interfaces.
%The more significantly interfaces deviate from the bulk, the easier the macroscopic theory fails as the size of the system shrinks.
The applicability of the macroscopic theory should be examined by studying the materials-specific 
dielectric screening properties of interfaces, one of motivations of this work.
%%}
The density functional theory (DFT) method fully takes the atomic structure of a device into account,
and microscopic electrostatics quantities such as potential and charge density can be solved self-consistently,
from which microscopic dielectric properties can be deduced.\cite{PhysRevLett.91.267601}
%%{\color{blue}
One can extract an effective dielectric constant from microscopic calculations to represent the dielectric screening of the interface,
which is not necessarily a constant but a quantity that depends on the thickness of the interface and other factors.
%This effective dielectric constant has its own physical significance to justify the applicability of the macroscopic theory of electrostatics.
%%}

The boundary condition is crucial in solving for the electrostatic potential.
In an FET device, for example,
the electrostatic potential at the surface of a metallic gate electrode should be constant,
and the electric field (derivative of the potential) in vacuum 
far away from the device should be zero.
This mixed boundary condition is different from the conventional and widely-implemented periodic boundary condition.
Otani {\etal}\cite{PhysRevB.73.115407} proposed a Green's-function-based 
effective screening medium (ESM) approach to solve the electrostatic potential
under several different boundary conditions, 
which is promising for simulating the electric-field effect in planar devices.

In this work, graphene-based FETT devices were simulated using a DFT+ESM method. 
The computational approach, based on first principles, enables us
to understand interface effects quantitatively, and thus enables computational design of functional systems. 
The rest of the paper is organized as follows.
In the next Section, details of the computational method are presented.
The distribution of free carriers and the band structures of graphene layers
under different gate voltages are presented in Sec.~\ref{sec:results}.
The effective dielectric constant of the \hbn{} thin barrier in FETT is analyzed in Sec.~\ref{sec:discuss}.
Finally, we give a summary on our results in Sec.~\ref{sec:summary}.

\section{simulation method}
\label{sec:method}

The structure of the FETT used in experiments\cite{Science.335.947} is schematically shown in Fig.~\ref{fig:struct}(a).
Doped silicon is used as the gate electrode, which is separated from the current-carrying \grt$|$\hbn$|$\grb\ trilayer channel
by insulating silicon oxide ($\sim300\,\nm$) and \hbn\ thin films ($\sim50\,\nm$).
The gate voltage is applied between the doped silicon and the trilayer.

In our DFT simulations, 
the doped silicon gate electrode was replaced by a semi-infinite ideally metallic medium for $z<-z_1$ 
[Fig.~\ref{fig:struct}(b); the $z$-direction is perpendicular to the \grt$|$\hbn$|$\grb{} trilayer], 
and the dielectrics between the gate electrode and  the \grt$|$\hbn$|$\grb{} trilayer were replaced by 
vacuum to save computational resources.
Thus the \grt$|$\hbn$|$\grb\ trilayer is sandwiched by a semi-infinite vacuum medium ($z>z_1$)
and an ideally metallic medium ($z<-z_1$), see Fig.~\ref{fig:struct}(b).
Note that the semi-infinite vacuum and metallic media are not explicitly included in calculations,
instead they are used as boundary conditions for the Hartree potential ($V_H$),
\begin{equation}
\label{eq:boundary}
{\partial V_H \over \partial z}|_{z=z_1} = 0, \qquad V_H|_{z=-z_1} = 0. 
\end{equation}
We adopted the ESM method\cite{PhysRevB.73.115407} to solve the Hartree potential.
The purpose to employ this method is twofold: 
(i) Long-range Coulomb interactions with periodic images are avoided.
(ii) The non-periodic boundary condition of Eq.~\eqref{eq:boundary} for the Hartree potential can be imposed.

A gate voltage can be simulated by adding extra electrons
or holes to the \grt$|$\hbn$|$\grb\ trilayer;\cite{PhysRevB.73.115407} 
the areal density of free carriers is proportional to the gate voltage.
In experiments the carrier density in graphene can be tuned
by gate voltage up to $\sim 10^{13}\, \mathrm{cm}^{-2}$,
which is equivalent to $5.24\times 10^{-3}$ electrons per primitive unit cell of graphene,
or an electric displacement field of $\Dg = 0.016\,\mathrm{C/m^2}$ in the dielectrics.

The measured interlayer distance between \hbn\ and graphene layers\cite{Nat.Mater.11.764} is $3.32\Ang$, and
the interlayer distance between \hbn\ layers $3.33\Ang$ is taken as half of the lattice constant $c$ of crystalline \hbn.\cite{BN-book}
In-plane lattice constants for both graphene and \hbn{} were set to $2.46\Ang$ due to the small lattice mismatch between them.
In the $x$-$y$ plane a periodic boundary condition was applied with a dense $155 \times 155$ Monkhorst-Pack\cite{PhysRevB.13.5188} $k$-mesh.
The cutoff energy for wave functions and the Methfessel-Paxton\cite{PhysRevB.40.3616} spreading energy
were taken to be $70\,\mathrm{Ryd}$ and $10^{-3}\,\mathrm{Ryd}$, respectively.
The Perdew-Burke-Ernzerhof parameterization\cite{PhysRevLett.77.3865} of the generalized gradient approximation (GGA)
to the exchange-correlation functional and 
Trouiller-Martins norm-conserving pseudopotentials were used.
DFT calculations were performed using the \textsc{Quantum ESPRESSO} package\cite{QE-2009}.

Charge density in the \grb$|$\hbn$|$\grt\ trilayer was calculated self-consistently for different gate voltages.
In order to illustrate the distribution of free carriers across the trilayer,
it is convenient to integrate the charge density in the $x$-$y$ plane
and define a charge density along $z$-direction $\rho(z)$.
The density of free carriers is defined as charge density difference
for a device under a finite gate voltage $\Vg$ with respect to that under zero gate voltage.
Compared to gate voltages, the electric displacement field $\Dg$ in the dielectrics
is a more convenient quantity, because it is independent of the thickness or the dielectric constant of the dielectrics.
Gate voltages are expressed as $\Dg$ hereafter.

\begin{figure}[b]
\begin{center}
\includegraphics[width=\linewidth]{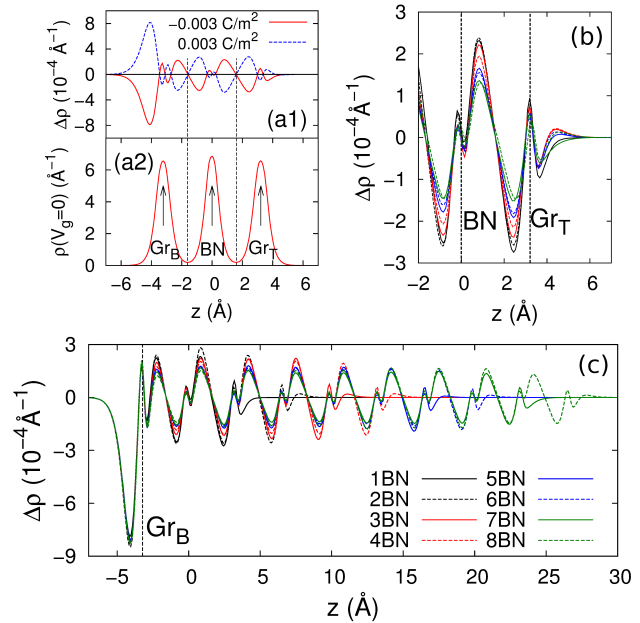}
\caption{
\label{fig:delta_rho}
(Color Online)
(a1) The distribution of free carriers under $\Dg=\pm 3\times 10^{-3}\,\mathrm{C/m^2}$ and
(a2) the charge density under zero gate voltage of a \grt$|$monolayer \hbn$|$\grb{}  system.
(b) The free carrier distribution in a graphene$|$multilayer \hbn$|$ graphene FETT
under $\Dg=3\times 10^{-3}\,\mathrm{C/m^2}$ 
with the positions of the top and (c) the bottom graphene layer aligned, respectively.
}
\end{center}
\end{figure}

\section{results}
\label{sec:results}

A FETT with a monolayer \hbn{} barrier is used as an example to illustrate the charge and free carrier densities 
in such devices..
The charge and free carrier density on each layer can be analyzed using the Bader decomposition,\cite{Bader} 
in which boundaries between atoms are defined by zero flux surfaces.
There are three peaks in the charge density curve $\rho(z)$,
corresponding to the \grt, \hbn, and \grb{} layers, respectively; cf. Fig.~\ref{fig:delta_rho}(a2).
Free carrier densities $\Delta\rho(z)$,
defined as the change in charge density induced by gate voltages,
under $\Dg=\pm 0.003\,\mathrm{C/m^2}$ are shown in Fig.~\ref{fig:delta_rho}(a1),
where it can be seen that they have opposite sign but almost the same amplitude,
in accordance with the electron-hole symmetry of graphene in the vicinity of the Fermi energy.
For the one-dimensional charge density $\rho(z)$, atomic layers are divided by minima of $\rho(z)$,
as denoted by vertical dashed lines in Fig.~\ref{fig:delta_rho}(a2)  
(zero flux surfaces shrink to points for the one-dimensional charge density.)
A large portion of the free carriers accumulate on \grb{}  at the side closer to the gate electrode; 
thus the free carrier density on \grb{}  is larger than that on \grt.
The asymmetric shape of $\Delta\rho(z)$ with respect to the central \hbn{}  layer is a consequence of
the asymmetric environment of the \grb$|$\hbn$|$\grt\ trilayer,
with the metallic gate electrode on one side and vacuum on the other side.

\begin{figure}[b]
\begin{center}
\includegraphics[width=\linewidth]{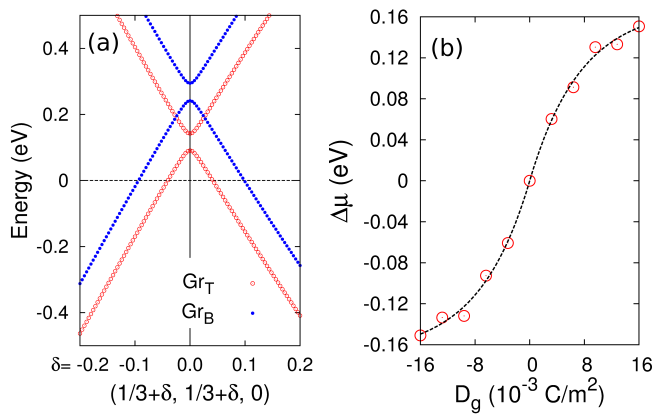}
\caption{
\label{fig:bandstructure}
(a) Band structure of a FETT with five layers of \hbn{}  as barrier at $\Dg=0.016\,\mathrm{C/m^2}$,
and (b) the difference in chemical potential between \grb{}  and \grt{}  as a function of gate voltage 
(the dashed line is a guide to the eye).
}
\end{center}
\end{figure}

The \hbn{}  barrier can screen the applied gate voltage by developing an internal electric polarization,
which reduces the effect of the gate voltage and the free carrier density in \grt.
We performed calculations on devices with a \hbn{}  barrier thickness of up to eight monolayers.
The free carrier density for these devices as a function of $z$ 
for $\Dg=3\times 10^{-3}\,\mathrm{C/m^2}$ 
is shown in Figs.~\ref{fig:delta_rho}(b) and \ref{fig:delta_rho}(c),
shifted to make the positions of the \grt{} (b) or \grb{}  (c) layers coincide.
The distribution of free carriers on \grb{} is almost independent of the thickness of the \hbn{} barrier, cf.  Fig.~\ref{fig:delta_rho}(c).
The amplitude of the electric polarization in the \hbn{}  barrier shrinks for thicker barriers (Fig.~\ref{fig:delta_rho}(c)),
and so does the free carrier density in \grt(Fig.~\ref{fig:delta_rho}(b)), indicating that thicker \hbn{}  barriers provide stronger screening.

The chemical potential of graphene, defined as the energy difference between the Fermi energy and the charge neutrality point,
can be efficiently tuned by using gate voltages because of the small DOS near the Fermi energy.%
\cite{Science.335.947,NNANO.8.100,NCOMMS.4.1794}
The band structure of a \grt$|$\hbn$|$\grb{} trilayer in the vicinity of the Fermi energy
has contributions only from   graphene layers,  because
\hbn{}  is a wide gap insulator with a calculated energy band gap of $4.77\eV$.
The band structure of graphene shows a tiny energy gap of $0.05\eV$ 
[Fig.~\ref{fig:bandstructure}(a)] induced by interaction with \hbn.\cite{PhysRevB.76.073103}
A FETT with a \hbn{}  barrier thicker than one monolayer shows no hybridization between \grb{}  and \grt{}  in its band structure.
As an example, the band structure of a FETT with a five-layer-thick \hbn{}  barrier for 
$\Dg=0.016\,\mathrm{C/m^2}$ is shown in Fig.~\ref{fig:bandstructure}(a), 
where bands from \grt{}  and \grb{}  shift upward rigidly by $\sim 0.1\eV$ and $\sim 0.2\eV$ with respect to the Fermi energy, respectively.
The bands from \grt{} are always shifted away from the Fermi level by a larger energy than those from \grb{}
[Fig.~\ref{fig:bandstructure}(b)].

\section{discussion}
\label{sec:discuss}

Macroscopic electrostatic models
have been used to calculate the electrostatic potential and free carrier density in graphene based FETT.
\cite{Science.335.947,JAP.113.136502,APL.101.033503,IEEE.60.268}
In these models, the \hbn\ barrier between graphene layers is assumed to have a dielectric constant equal to that of the \hbn\ crystal.
However, interfaces can exhibit significantly different dielectric properties compared to the bulk,\cite{PhysRevLett.91.267601}
and the dielectric properties of a few-layer-thick \hbn{}  barrier sandwiched between two graphene layers need to be revisited.
We seek to find the (effective) dielectric constant $\eBN$ of a \hbn{}  barrier in a FETT and compare to its bulk value,
in order to investigate if the dielectric constant is still a valid physical concept for thin \hbn{}  films
and to investigate whether $\eBN$ of a \hbn{}  barrier is modified by the interfaces with graphene layers.

\begin{figure}[b!]
\begin{center}
\includegraphics[width=\linewidth]{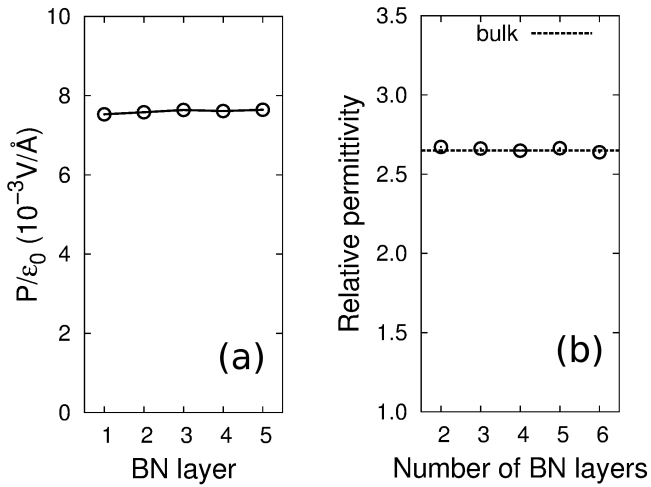}
\caption{
\label{fig:dielectric_GrBNGr}
(a) The polarization of \hbn{} layers in a FETT with a five-layer \hbn{} thin film
induced by an gate voltage of $\Dg=6.114 \times 10^{-3}\,\mathrm{C/m^2}$.
(b) The calculated average relative dielectric permittivity of \hbn{} thin films
embedded by graphene layers as a function of the number of \hbn{} atomic layers,
and the dashed line denotes the relative dielectric permittivity of bulk \hbn{}.
}
\end{center}
\end{figure}

%{\color{blue}  %%%% 2015 April
The dielectric constants of \hbn{} thin films ($\eBN$) sandwiched between graphene layers
can be deduced from the electric field inside the \hbn{} thin film ($\Eint$)
and the electric polarization ($P$) induced by gate voltages,
\begin{equation}
\label{eq:epsilon1}
\eBN = \frac{\Eint+P/\epsilon_0}{\Eint}, 
\end{equation}
where $\Eint$ is equal to the slope of the self-consistent Kohn-Sham effective potential in the $z$-direction.
$\Eint$ is also related to the difference between chemical potentials of graphene layers $\Delta\mu$ by
$ e \Eint d =\Delta\mu $, where $d$ is the distance between the two graphene layers.

The polarization can be expressed as the summation of centers of Wannier functions
according to the modern theory of polarization.
\cite{PhysRevB.47.1651,RevModPhys.66.899}
%\cite{PhysRevLett.91.267601,PhysRevB.71.144104}
In practice we followed the method proposed in Refs.~\onlinecite{PhysRevB.75.205121,PhysRevB.80.224110}
which extends the Wannier function theory of polarization
\cite{PhysRevB.47.1651,RevModPhys.66.899,PhysRevB.71.144104,PhysRevLett.97.107602}
to metal-insulator heterostructures.

The hybrid Wannier functions
\cite{PhysRevB.56.12847,PhysRevB.64.115202},
which are exponentially localized in the $z$ direction
but Bloch-like in the $x$-$y$ plane, 
were constructed using the parallel-transport method.\cite{PhysRevB.56.12847,PhysRevB.64.115202}
The first Brillouin zone was sampled by discrete $k$-points of the type
$\mathbf{k} = \mathbf{k_\perp} + j\,\mathbf{b}$, 
where the vectors $\mathbf{k_\perp}$ form a
$ N_\perp \times N_\perp$ uniform mesh in the $x$-$y$ plane, 
and $ \mathbf{b} = (0,0,2\pi N_\| /L )$ is along the $z$-direction with $L$ the height of the unit cell
and $N_\|$ the number of $k$-points along the $z$-direction.
The total number of $k$-points is $N_\perp^2 N_\|$.
We used $N_\perp=7$ and $N_\|=3$, which are sufficient to converge the polarization.

The matrices
$ M_{mn}(\mathbf{k} ) = \braket{ u_{m,\mathbf{k}} }{ u_{n,\mathbf{k} + \mathbf{b}}}$ 
were constructed where $\ket{ u_{m,\mathbf{k}} }$ is periodic under lattice translations and $n$ is the band index; 
$\ket{ u_{m,\mathbf{k}} }$ is normalized such that
$\braket{ u_{m,\mathbf{k}}  } { u_{m,\mathbf{k}} } =1$.
Singular value decomposition of each $M$ matrix was done utilizing the LAPACK library: $M=U \Sigma V^{\dagger}$
where $U$ and $V$ are complex unitary matrices and $\Sigma$ is a diagonal matrix with diagonal elements very close to 1 for small $\mathbf{b}$.
A new matrix $\tilde{M} = UV^{\dagger}$ was constructed corresponding to each $M$ matrix, and
a global matrix $\Lambda(\mathbf{k}_{\perp}) = \prod_{j=0}^{N_\|-1} \tilde{M}( \mathbf{k_\perp} + j\mathbf{b}) $
was then constructed for each $\mathbf{k_\perp}$ point.
The centers of hybrid Wannier functions were calculated as $z_m = (-L/2\pi) \,\textrm{Im}[ \ln \lambda_m]$,
where the $\lambda_m$ are the eigenvalues of $\Lambda$.

The number of bands considered to construct the hybrid Wannier functions,
i.e., the dimension of the $M$ matrices, is equal to four for each graphene or \hbn{} atomic layer,
the number of occupied bands in most of the first Brillouin zone
except for the small portion near the $K$-point (see Fig.~\ref{fig:bandstructure}).
This choice does not affect the calculations of the polarization inside \hbn{} thin films
because the bands near the Fermi energy are contributed by graphene layers.

Four of the resulting hybrid Wannier functions can be assigned to each graphene or \hbn{} layer
according to the positions of their centers.
Two of them are very close to the atomic plane  (within $10^{-3} \Ang$)  and
the other two are located about $ 0.4 \Ang $  above and beneath the atomic plane respectively.
The center of charge for each atomic layer is equal to the average value of the corresponding four Wannier functions.
The dipole moment corresponding to each \hbn{} layer was calculated using the shift of the center of charge under an electric field,
and the polarization was calculated with the thickness of each \hbn{} layer set to be $ 3.33\,\Ang $.

The calculated polarization of the \hbn{} layer adjacent to the graphene layers
is almost the same as  \hbn{} layers deeply inside [see Fig.~\ref{fig:dielectric_GrBNGr}(a)],
indicating that interface with graphene layers has little effect on the dielectric properties of \hbn{} thin films.
As a result, the average dielectric constant for \hbn{} thin films embedded by graphene layers
is independent of the thickness; see Fig.~\ref{fig:dielectric_GrBNGr}(b).

%} %color blue %%%% 2015 April

\begin{figure}[b!]
\begin{center}
\includegraphics[width=\linewidth]{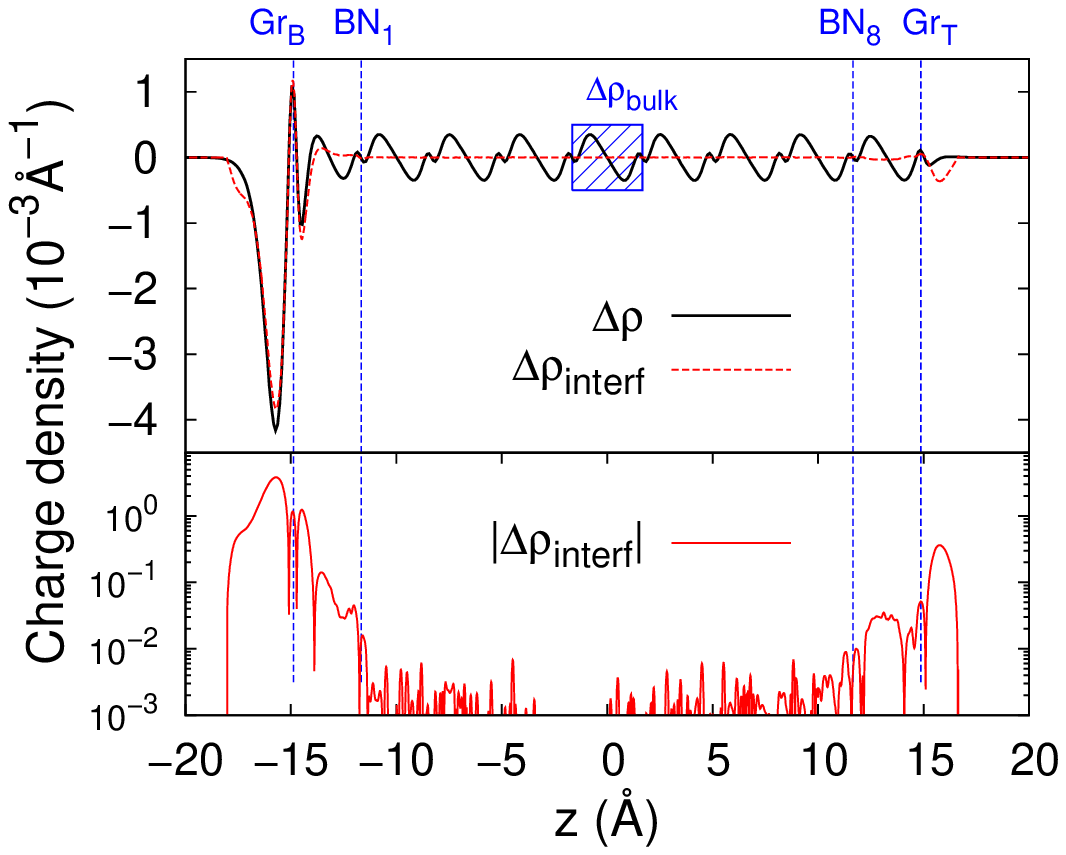}
\caption{
\label{fig:surface_charge}
(Color online)
Interface charge redistribution \rhoif{}  in a FETT device with eight-layer \hbn{} for  
$\Dg=0.016\,\mathrm{C/m^2}$, 
where \rhoif{}  is  the total charge redistribution $\Delta\rho$ 
minus the bulk charge redistribution $\rhob$ at the center of \hbn; 
\rhob{}  is denoted by the patterned rectangle in the top panel.
The absolute value of \rhoif{}  is plotted on a logarithmic scale in the bottom panel.
The positions of the atomic planes of \grb, \grt{}  and the two interface \hbn{} layers  
$\mathrm{BN_1}$ and $\mathrm{BN_8}$ are denoted by vertical dashed lines.
}
\end{center}
\end{figure}

The effect of an interface with graphene on the dielectric properties of \hbn{} 
can be analyzed using the interface charge redistribution, denoted as \rhoif{}
 and defined as the difference in charge redistribution
at the interface with respect to that in the bulk (denoted as \rhob).
In practice we used the charge redistribution at the center of the \hbn{}  as the bulk,
as shown in Fig.~\ref{fig:surface_charge}.
Thus one can obtain \rhoif{}  by subtracting \rhob{}  from the total charge redistribution.
Because \hbn{}  is a wide-band-gap insulator, \rhoif{}  should decay exponentially away from the interface.
The strength of the interface effect is determined by the amplitude of \rhoif{}  near the interface.

An example the charge redistribution in a FETT device with eight-layer \hbn{} for $\Dg=0.016\,\mathrm{C/m^2}$
is shown in Fig.~\ref{fig:surface_charge}.
The difference \rhoif{}  is large near the graphene layers, decays quickly into \hbn, 
and becomes invisible after crossing the first \hbn\ atomic layers
($\mathrm{BN_1}$ and $\mathrm{BN_8}$ in Fig.~\ref{fig:surface_charge}).
The amplitude of \rhoif{}  is also presented on a logarithmic scale
in the lower panel of Fig.~\ref{fig:surface_charge}.
The decay of \rhoif{}  into \hbn{}  is approximately exponential.
Most importantly, the amplitude of \rhoif{}  is about $50$ times smaller than that of \rhob{} 
between $\mathrm{BN_1}$ and $\mathrm{BN_8}$ in Fig.~\ref{fig:surface_charge}: 
\rhoif{}  is very small inside the atomic plane of the first \hbn{}  layer.
As a result, any interface effect on the dielectric properties of \hbn\ is very weak,
which explains why the calculated dielectric constant of \hbn{}  in FETT devices is close to the bulk value.

%{\color{blue}  %%% 2015 April

\begin{figure}[h]
\begin{center}
\includegraphics[width=\linewidth]{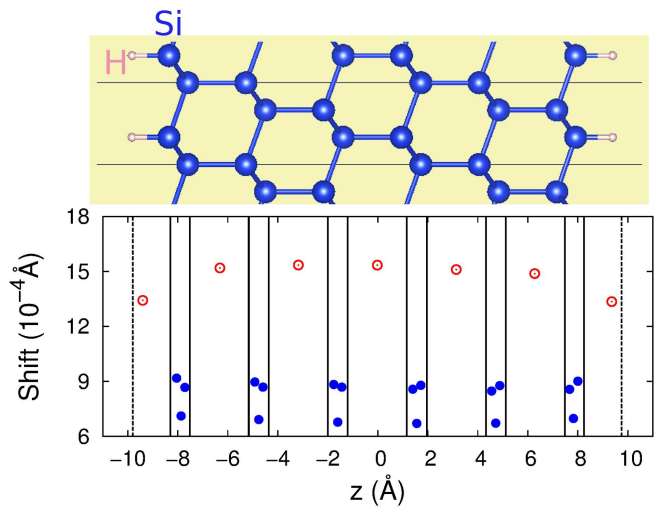}
\caption{
\label{fig:Si111}
(Color online)
(Upper panel)
The atomic structure of a hydrogen saturated Si(111) thin film with a thickness of about $ 2\,\nm $.
Si and H atoms are represented by large blue and small pink spheres, respectively.
Solid lines denote the boundary of the unit cell.
(Lower panel)
The shift of the hybrid Wannier functions induced by an electric field
of $ 0.0385\,\mathrm{V/\Ang} $ along the $z$-direction.
Red circles ($\circ$) and blue discs ($\bullet$) represent
hybrid Wannier functions with higher and lower polarizability, respectively.
}
\end{center}
\end{figure}

We also compared the \hbn{} thin films with silicon thin films, 
because the latter are known to exhibit lower dielectric permittivity than for the corresponding bulk.
\cite{PhysRevLett.91.267601,PhysRevB.71.144104}
The in-plane lattice constant of Si(111) slabs was chosen to be the experimental lattice constant of {\it fcc}-Si.
Dangling bonds on both surfaces are saturated by hydrogen, as shown in the upper panel of Fig.~\ref{fig:Si111}, 
and the Si-H bond length is $ 1.50\,\Ang $ after structure optimization.

We plotted in Fig.~\ref{fig:Si111} the shift of each hybrid Wannier functions
induced by an electric field of $ 0.0385\,\mathrm{V/\Ang} $ along the $z$-direction,
where the electric field was applied using the ESM method using a metal$|$slab$|$metal configuration.
The hybrid Wannier functions can be divided into two categories according to their polarizability.
The hybrid Wannier functions located at canted Si-Si bonds with respect to the $z$-direction
(denoted as $\bullet$ in Fig.~\ref{fig:Si111}) exhibit lower polarizability, 
and they show negligible deviations at the surface.
On the other hand, the hybrid Wannier functions located at parallel Si-Si or Si-H bonds
with respect to the $z$-direction (denoted as $\circ$ in Fig.~\ref{fig:Si111}) exhibit higher polarizability.
We also observed that the hybrid Wannier functions of Si-H bonds at the surface show a polarizability 12\%   lower 
than those corresponding to parallel Si-Si bonds.
The 12\%  lower polarizability of the Si(111) surfaces shown in Fig.~\ref{fig:Si111}
is not as severe as reported by previous studies.\cite{PhysRevB.71.144104,JAP.99.054309}
In those studies, % \cite{PhysRevB.71.144104,JAP.99.054309},
the macroscopic polarization and dielectric constant were obtained after a smoothing procedure.
The purpose of the smoothing procedure is to eliminate the dielectric nonlocality, 
but this procedure reduces the dielectric permittivity of surfaces artificially
because the dielectric constant of vacuum is smeared into the surface.

%} %%% color blue

\section{summary}
\label{sec:summary}

The distribution of free carriers and the band structure of graphene layers
in graphene based FETT have been simulated using the DFT+ESM method.
The dielectric properties of \hbn{}  thin films sandwiched between graphene layers in FETT
were investigated using the theory of microscopic permittivity 
and found to have a dielectric permittivity close to that of crystalline \hbn.
The small amplitude of interface charge redistribution inside the atomic plane of the first \hbn{}  layer
proves that the effect of the interface with graphene on the dielectric properties of \hbn{}  is weak.

In this study we have demonstrated the DFT+ESM method as a promising approach
to simulate field-effect devices with a planar structure.
Once the charge density and effective potential of a field-effect device
are self-consistently obtained, the scattering of transport electrons
and electric conductivity can be calculated using scattering theory.

\begin{acknowledgments}

This work was supported by the US Department of Energy (DOE),
Office of Basic Energy Sciences (BES),
under Contract No. DE-FG02-02ER45995.
This research used resources of the National Energy Research Scientific Computing Center.

\end{acknowledgments}

%%%%%%%%%%%%%%%%%%%%%%%%%%%%%%%%%%%%%%
%
%\bibliography{../../bib-collection}

\begin{thebibliography}{32}%
\makeatletter
\providecommand \@ifxundefined [1]{%
 \@ifx{#1\undefined}
}%
\providecommand \@ifnum [1]{%
 \ifnum #1\expandafter \@firstoftwo
 \else \expandafter \@secondoftwo
 \fi
}%
\providecommand \@ifx [1]{%
 \ifx #1\expandafter \@firstoftwo
 \else \expandafter \@secondoftwo
 \fi
}%
\providecommand \natexlab [1]{#1}%
\providecommand \enquote  [1]{``#1''}%
\providecommand \bibnamefont  [1]{#1}%
\providecommand \bibfnamefont [1]{#1}%
\providecommand \citenamefont [1]{#1}%
\providecommand \href@noop [0]{\@secondoftwo}%
\providecommand \href [0]{\begingroup \@sanitize@url \@href}%
\providecommand \@href[1]{\@@startlink{#1}\@@href}%
\providecommand \@@href[1]{\endgroup#1\@@endlink}%
\providecommand \@sanitize@url [0]{\catcode `\\12\catcode `\$12\catcode
  `\&12\catcode `\#12\catcode `\^12\catcode `\_12\catcode `\%12\relax}%
\providecommand \@@startlink[1]{}%
\providecommand \@@endlink[0]{}%
\providecommand \url  [0]{\begingroup\@sanitize@url \@url }%
\providecommand \@url [1]{\endgroup\@href {#1}{\urlprefix }}%
\providecommand \urlprefix  [0]{URL }%
\providecommand \Eprint [0]{\href }%
\providecommand \doibase [0]{http://dx.doi.org/}%
\providecommand \selectlanguage [0]{\@gobble}%
\providecommand \bibinfo  [0]{\@secondoftwo}%
\providecommand \bibfield  [0]{\@secondoftwo}%
\providecommand \translation [1]{[#1]}%
\providecommand \BibitemOpen [0]{}%
\providecommand \bibitemStop [0]{}%
\providecommand \bibitemNoStop [0]{.\EOS\space}%
\providecommand \EOS [0]{\spacefactor3000\relax}%
\providecommand \BibitemShut  [1]{\csname bibitem#1\endcsname}%
\let\auto@bib@innerbib\@empty
%</preamble>
\bibitem [{\citenamefont {Novoselov}\ \emph {et~al.}(2004)\citenamefont
  {Novoselov}, \citenamefont {Geim}, \citenamefont {Morozov}, \citenamefont
  {Jiang}, \citenamefont {Zhang}, \citenamefont {Dubonos}, \citenamefont
  {Grigorieva},\ and\ \citenamefont {Firsov}}]{Science-graphene}%
  \BibitemOpen
  \bibfield  {author} {\bibinfo {author} {\bibfnamefont {K.~S.}\ \bibnamefont
  {Novoselov}}, \bibinfo {author} {\bibfnamefont {A.~K.}\ \bibnamefont {Geim}},
  \bibinfo {author} {\bibfnamefont {S.~V.}\ \bibnamefont {Morozov}}, \bibinfo
  {author} {\bibfnamefont {D.}~\bibnamefont {Jiang}}, \bibinfo {author}
  {\bibfnamefont {Y.}~\bibnamefont {Zhang}}, \bibinfo {author} {\bibfnamefont
  {S.~V.}\ \bibnamefont {Dubonos}}, \bibinfo {author} {\bibfnamefont {I.~V.}\
  \bibnamefont {Grigorieva}}, \ and\ \bibinfo {author} {\bibfnamefont {A.~A.}\
  \bibnamefont {Firsov}},\ }\href {\doibase 10.1126/science.1102896} {\bibfield
   {journal} {\bibinfo  {journal} {Science}\ }\textbf {\bibinfo {volume}
  {306}},\ \bibinfo {pages} {666} (\bibinfo {year} {2004})}\BibitemShut
  {NoStop}%
\bibitem [{\citenamefont {Castro}\ \emph {et~al.}(2007)\citenamefont {Castro},
  \citenamefont {Novoselov}, \citenamefont {Morozov}, \citenamefont {Peres},
  \citenamefont {dos Santos}, \citenamefont {Nilsson}, \citenamefont {Guinea},
  \citenamefont {Geim},\ and\ \citenamefont {Neto}}]{PhysRevLett.99.216802}%
  \BibitemOpen
  \bibfield  {author} {\bibinfo {author} {\bibfnamefont {E.~V.}\ \bibnamefont
  {Castro}}, \bibinfo {author} {\bibfnamefont {K.~S.}\ \bibnamefont
  {Novoselov}}, \bibinfo {author} {\bibfnamefont {S.~V.}\ \bibnamefont
  {Morozov}}, \bibinfo {author} {\bibfnamefont {N.~M.~R.}\ \bibnamefont
  {Peres}}, \bibinfo {author} {\bibfnamefont {J.~M. B.~L.}\ \bibnamefont {dos
  Santos}}, \bibinfo {author} {\bibfnamefont {J.}~\bibnamefont {Nilsson}},
  \bibinfo {author} {\bibfnamefont {F.}~\bibnamefont {Guinea}}, \bibinfo
  {author} {\bibfnamefont {A.~K.}\ \bibnamefont {Geim}}, \ and\ \bibinfo
  {author} {\bibfnamefont {A.~H.~C.}\ \bibnamefont {Neto}},\ }\href {\doibase
  10.1103/PhysRevLett.99.216802} {\bibfield  {journal} {\bibinfo  {journal}
  {Phys. Rev. Lett.}\ }\textbf {\bibinfo {volume} {99}},\ \bibinfo {pages}
  {216802} (\bibinfo {year} {2007})}\BibitemShut {NoStop}%
\bibitem [{\citenamefont {Oostinga}\ \emph {et~al.}(2007)\citenamefont
  {Oostinga}, \citenamefont {Heersche}, \citenamefont {Liu}, \citenamefont
  {Morpurgo},\ and\ \citenamefont {Vandersypen}}]{Nat.Mater.7.151}%
  \BibitemOpen
  \bibfield  {author} {\bibinfo {author} {\bibfnamefont {J.~B.}\ \bibnamefont
  {Oostinga}}, \bibinfo {author} {\bibfnamefont {H.~B.}\ \bibnamefont
  {Heersche}}, \bibinfo {author} {\bibfnamefont {X.}~\bibnamefont {Liu}},
  \bibinfo {author} {\bibfnamefont {A.~F.}\ \bibnamefont {Morpurgo}}, \ and\
  \bibinfo {author} {\bibfnamefont {L.~M.~K.}\ \bibnamefont {Vandersypen}},\
  }\href@noop {} {\bibfield  {journal} {\bibinfo  {journal} {Nat. Mater.}\
  }\textbf {\bibinfo {volume} {7}},\ \bibinfo {pages} {151} (\bibinfo {year}
  {2007})}\BibitemShut {NoStop}%
\bibitem [{\citenamefont {Han}\ \emph {et~al.}(2007)\citenamefont {Han},
  \citenamefont {\"Ozyilmaz}, \citenamefont {Zhang},\ and\ \citenamefont
  {Kim}}]{PhysRevLett.98.206805}%
  \BibitemOpen
  \bibfield  {author} {\bibinfo {author} {\bibfnamefont {M.~Y.}\ \bibnamefont
  {Han}}, \bibinfo {author} {\bibfnamefont {B.}~\bibnamefont {\"Ozyilmaz}},
  \bibinfo {author} {\bibfnamefont {Y.}~\bibnamefont {Zhang}}, \ and\ \bibinfo
  {author} {\bibfnamefont {P.}~\bibnamefont {Kim}},\ }\href {\doibase
  10.1103/PhysRevLett.98.206805} {\bibfield  {journal} {\bibinfo  {journal}
  {Phys. Rev. Lett.}\ }\textbf {\bibinfo {volume} {98}},\ \bibinfo {pages}
  {206805} (\bibinfo {year} {2007})}\BibitemShut {NoStop}%
\bibitem [{\citenamefont {Elias}\ \emph {et~al.}(2009)\citenamefont {Elias},
  \citenamefont {Nair}, \citenamefont {Mohiuddin}, \citenamefont {Morozov},
  \citenamefont {Blake}, \citenamefont {Halsall}, \citenamefont {Ferrari},
  \citenamefont {Boukhvalov}, \citenamefont {Katsnelson}, \citenamefont
  {Geim},\ and\ \citenamefont {Novoselov}}]{Science.323.610}%
  \BibitemOpen
  \bibfield  {author} {\bibinfo {author} {\bibfnamefont {D.~C.}\ \bibnamefont
  {Elias}}, \bibinfo {author} {\bibfnamefont {R.~R.}\ \bibnamefont {Nair}},
  \bibinfo {author} {\bibfnamefont {T.~M.~G.}\ \bibnamefont {Mohiuddin}},
  \bibinfo {author} {\bibfnamefont {S.~V.}\ \bibnamefont {Morozov}}, \bibinfo
  {author} {\bibfnamefont {P.}~\bibnamefont {Blake}}, \bibinfo {author}
  {\bibfnamefont {M.~P.}\ \bibnamefont {Halsall}}, \bibinfo {author}
  {\bibfnamefont {A.~C.}\ \bibnamefont {Ferrari}}, \bibinfo {author}
  {\bibfnamefont {D.~W.}\ \bibnamefont {Boukhvalov}}, \bibinfo {author}
  {\bibfnamefont {M.~I.}\ \bibnamefont {Katsnelson}}, \bibinfo {author}
  {\bibfnamefont {A.~K.}\ \bibnamefont {Geim}}, \ and\ \bibinfo {author}
  {\bibfnamefont {K.~S.}\ \bibnamefont {Novoselov}},\ }\href {\doibase
  10.1126/science.1167130} {\bibfield  {journal} {\bibinfo  {journal}
  {Science}\ }\textbf {\bibinfo {volume} {323}},\ \bibinfo {pages} {610}
  (\bibinfo {year} {2009})}\BibitemShut {NoStop}%
\bibitem [{\citenamefont {Britnell}\ \emph {et~al.}(2012)\citenamefont
  {Britnell}, \citenamefont {Gorbachev}, \citenamefont {Jalil}, \citenamefont
  {Belle}, \citenamefont {Schedin}, \citenamefont {Mishchenko}, \citenamefont
  {Georgiou}, \citenamefont {Katsnelson}, \citenamefont {Eaves}, \citenamefont
  {Morozov}, \citenamefont {Peres}, \citenamefont {Leist}, \citenamefont
  {Geim}, \citenamefont {Novoselov},\ and\ \citenamefont
  {Ponomarenko}}]{Science.335.947}%
  \BibitemOpen
  \bibfield  {author} {\bibinfo {author} {\bibfnamefont {L.}~\bibnamefont
  {Britnell}}, \bibinfo {author} {\bibfnamefont {R.~V.}\ \bibnamefont
  {Gorbachev}}, \bibinfo {author} {\bibfnamefont {R.}~\bibnamefont {Jalil}},
  \bibinfo {author} {\bibfnamefont {B.~D.}\ \bibnamefont {Belle}}, \bibinfo
  {author} {\bibfnamefont {F.}~\bibnamefont {Schedin}}, \bibinfo {author}
  {\bibfnamefont {A.}~\bibnamefont {Mishchenko}}, \bibinfo {author}
  {\bibfnamefont {T.}~\bibnamefont {Georgiou}}, \bibinfo {author}
  {\bibfnamefont {M.~I.}\ \bibnamefont {Katsnelson}}, \bibinfo {author}
  {\bibfnamefont {L.}~\bibnamefont {Eaves}}, \bibinfo {author} {\bibfnamefont
  {S.~V.}\ \bibnamefont {Morozov}}, \bibinfo {author} {\bibfnamefont
  {N.~M.~R.}\ \bibnamefont {Peres}}, \bibinfo {author} {\bibfnamefont
  {J.}~\bibnamefont {Leist}}, \bibinfo {author} {\bibfnamefont {A.~K.}\
  \bibnamefont {Geim}}, \bibinfo {author} {\bibfnamefont {K.~S.}\ \bibnamefont
  {Novoselov}}, \ and\ \bibinfo {author} {\bibfnamefont {L.~A.}\ \bibnamefont
  {Ponomarenko}},\ }\href {\doibase 10.1126/science.1218461} {\bibfield
  {journal} {\bibinfo  {journal} {Science}\ }\textbf {\bibinfo {volume}
  {335}},\ \bibinfo {pages} {947} (\bibinfo {year} {2012})}\BibitemShut
  {NoStop}%
\bibitem [{\citenamefont {Georgiou}\ \emph {et~al.}(2013)\citenamefont
  {Georgiou}, \citenamefont {Jalil}, \citenamefont {Belle}, \citenamefont
  {Britnell}, \citenamefont {Gorbachev}, \citenamefont {Morozov}, \citenamefont
  {Kim}, \citenamefont {Gholinia}, \citenamefont {Haigh}, \citenamefont
  {Makarovsky}, \citenamefont {Eaves}, \citenamefont {Ponomarenko},
  \citenamefont {Geim}, \citenamefont {Novoselov},\ and\ \citenamefont
  {Mishchenko}}]{NNANO.8.100}%
  \BibitemOpen
  \bibfield  {author} {\bibinfo {author} {\bibfnamefont {T.}~\bibnamefont
  {Georgiou}}, \bibinfo {author} {\bibfnamefont {R.}~\bibnamefont {Jalil}},
  \bibinfo {author} {\bibfnamefont {B.~D.}\ \bibnamefont {Belle}}, \bibinfo
  {author} {\bibfnamefont {L.}~\bibnamefont {Britnell}}, \bibinfo {author}
  {\bibfnamefont {R.~V.}\ \bibnamefont {Gorbachev}}, \bibinfo {author}
  {\bibfnamefont {S.~V.}\ \bibnamefont {Morozov}}, \bibinfo {author}
  {\bibfnamefont {Y.-J.}\ \bibnamefont {Kim}}, \bibinfo {author} {\bibfnamefont
  {A.}~\bibnamefont {Gholinia}}, \bibinfo {author} {\bibfnamefont {S.~J.}\
  \bibnamefont {Haigh}}, \bibinfo {author} {\bibfnamefont {O.}~\bibnamefont
  {Makarovsky}}, \bibinfo {author} {\bibfnamefont {L.}~\bibnamefont {Eaves}},
  \bibinfo {author} {\bibfnamefont {L.~A.}\ \bibnamefont {Ponomarenko}},
  \bibinfo {author} {\bibfnamefont {A.~K.}\ \bibnamefont {Geim}}, \bibinfo
  {author} {\bibfnamefont {K.~S.}\ \bibnamefont {Novoselov}}, \ and\ \bibinfo
  {author} {\bibfnamefont {A.}~\bibnamefont {Mishchenko}},\ }\href {\doibase
  10.1038/NNANO.2012.224} {\bibfield  {journal} {\bibinfo  {journal} {Nature
  Nanotech.}\ }\textbf {\bibinfo {volume} {8}},\ \bibinfo {pages} {100}
  (\bibinfo {year} {2013})}\BibitemShut {NoStop}%
\bibitem [{\citenamefont {Britnell}\ \emph {et~al.}(2013)\citenamefont
  {Britnell}, \citenamefont {Gorbachev}, \citenamefont {Geim}, \citenamefont
  {Ponomarenko}, \citenamefont {Mishchenko}, \citenamefont {Greenaway},
  \citenamefont {Fromhold}, \citenamefont {Novoselov},\ and\ \citenamefont
  {Eaves}}]{NCOMMS.4.1794}%
  \BibitemOpen
  \bibfield  {author} {\bibinfo {author} {\bibfnamefont {L.}~\bibnamefont
  {Britnell}}, \bibinfo {author} {\bibfnamefont {R.~V.}\ \bibnamefont
  {Gorbachev}}, \bibinfo {author} {\bibfnamefont {A.~K.}\ \bibnamefont {Geim}},
  \bibinfo {author} {\bibfnamefont {L.~A.}\ \bibnamefont {Ponomarenko}},
  \bibinfo {author} {\bibfnamefont {A.}~\bibnamefont {Mishchenko}}, \bibinfo
  {author} {\bibfnamefont {M.~T.}\ \bibnamefont {Greenaway}}, \bibinfo {author}
  {\bibfnamefont {T.~M.}\ \bibnamefont {Fromhold}}, \bibinfo {author}
  {\bibfnamefont {K.~S.}\ \bibnamefont {Novoselov}}, \ and\ \bibinfo {author}
  {\bibfnamefont {L.}~\bibnamefont {Eaves}},\ }\href {\doibase
  10.1038/ncomms2817} {\bibfield  {journal} {\bibinfo  {journal} {Nature
  Commun.}\ }\textbf {\bibinfo {volume} {4}},\ \bibinfo {pages} {1794}
  (\bibinfo {year} {2013})}\BibitemShut {NoStop}%
\bibitem [{\citenamefont {Luo}\ \emph {et~al.}(2007)\citenamefont {Luo},
  \citenamefont {Li}, \citenamefont {Xia},\ and\ \citenamefont
  {Wang}}]{APL.90.143108}%
  \BibitemOpen
  \bibfield  {author} {\bibinfo {author} {\bibfnamefont {J.-W.}\ \bibnamefont
  {Luo}}, \bibinfo {author} {\bibfnamefont {S.-S.}\ \bibnamefont {Li}},
  \bibinfo {author} {\bibfnamefont {J.-B.}\ \bibnamefont {Xia}}, \ and\
  \bibinfo {author} {\bibfnamefont {L.-W.}\ \bibnamefont {Wang}},\ }\href@noop
  {} {\bibfield  {journal} {\bibinfo  {journal} {Appl. Phys. Lett.}\ }\textbf
  {\bibinfo {volume} {90}},\ \bibinfo {pages} {143108} (\bibinfo {year}
  {2007})}\BibitemShut {NoStop}%
\bibitem [{\citenamefont {Jackson}(1998)}]{Jackson}%
  \BibitemOpen
  \bibfield  {author} {\bibinfo {author} {\bibfnamefont {J.~D.}\ \bibnamefont
  {Jackson}},\ }\href@noop {} {\emph {\bibinfo {title} {Classical
  Electrodynamics}}},\ \bibinfo {edition} {3rd}\ ed.\ (\bibinfo  {publisher}
  {Wiley},\ \bibinfo {year} {1998})\BibitemShut {NoStop}%
\bibitem [{\citenamefont {Giustino}\ \emph {et~al.}(2003)\citenamefont
  {Giustino}, \citenamefont {Umari},\ and\ \citenamefont
  {Pasquarello}}]{PhysRevLett.91.267601}%
  \BibitemOpen
  \bibfield  {author} {\bibinfo {author} {\bibfnamefont {F.}~\bibnamefont
  {Giustino}}, \bibinfo {author} {\bibfnamefont {P.}~\bibnamefont {Umari}}, \
  and\ \bibinfo {author} {\bibfnamefont {A.}~\bibnamefont {Pasquarello}},\
  }\href {\doibase 10.1103/PhysRevLett.91.267601} {\bibfield  {journal}
  {\bibinfo  {journal} {Phys. Rev. Lett.}\ }\textbf {\bibinfo {volume} {91}},\
  \bibinfo {pages} {267601} (\bibinfo {year} {2003})}\BibitemShut {NoStop}%
\bibitem [{\citenamefont {Otani}\ and\ \citenamefont
  {Sugino}(2006)}]{PhysRevB.73.115407}%
  \BibitemOpen
  \bibfield  {author} {\bibinfo {author} {\bibfnamefont {M.}~\bibnamefont
  {Otani}}\ and\ \bibinfo {author} {\bibfnamefont {O.}~\bibnamefont {Sugino}},\
  }\href {\doibase 10.1103/PhysRevB.73.115407} {\bibfield  {journal} {\bibinfo
  {journal} {Phys. Rev. B}\ }\textbf {\bibinfo {volume} {73}},\ \bibinfo
  {pages} {115407} (\bibinfo {year} {2006})}\BibitemShut {NoStop}%
\bibitem [{\citenamefont {Haigh}\ \emph {et~al.}(2012)\citenamefont {Haigh},
  \citenamefont {Gholinia}, \citenamefont {Jalil}, \citenamefont {Romani},
  \citenamefont {Britnell}, \citenamefont {Elias}, \citenamefont {Novoselov},
  \citenamefont {Ponomarenko}, \citenamefont {Geim},\ and\ \citenamefont
  {Gorbachev}}]{Nat.Mater.11.764}%
  \BibitemOpen
  \bibfield  {author} {\bibinfo {author} {\bibfnamefont {S.~J.}\ \bibnamefont
  {Haigh}}, \bibinfo {author} {\bibfnamefont {A.}~\bibnamefont {Gholinia}},
  \bibinfo {author} {\bibfnamefont {R.}~\bibnamefont {Jalil}}, \bibinfo
  {author} {\bibfnamefont {S.}~\bibnamefont {Romani}}, \bibinfo {author}
  {\bibfnamefont {L.}~\bibnamefont {Britnell}}, \bibinfo {author}
  {\bibfnamefont {D.~C.}\ \bibnamefont {Elias}}, \bibinfo {author}
  {\bibfnamefont {K.~S.}\ \bibnamefont {Novoselov}}, \bibinfo {author}
  {\bibfnamefont {L.~A.}\ \bibnamefont {Ponomarenko}}, \bibinfo {author}
  {\bibfnamefont {A.~K.}\ \bibnamefont {Geim}}, \ and\ \bibinfo {author}
  {\bibfnamefont {R.}~\bibnamefont {Gorbachev}},\ }\href@noop {} {\bibfield
  {journal} {\bibinfo  {journal} {Nat. Mater.}\ }\textbf {\bibinfo {volume}
  {11}},\ \bibinfo {pages} {764} (\bibinfo {year} {2012})}\BibitemShut
  {NoStop}%
\bibitem [{\citenamefont {Levinshtein}\ \emph {et~al.}(2001)\citenamefont
  {Levinshtein}, \citenamefont {Rumyantsev},\ and\ \citenamefont
  {Shur}}]{BN-book}%
  \BibitemOpen
  \bibinfo {editor} {\bibfnamefont {M.~E.}\ \bibnamefont {Levinshtein}},
  \bibinfo {editor} {\bibfnamefont {S.~L.}\ \bibnamefont {Rumyantsev}}, \ and\
  \bibinfo {editor} {\bibfnamefont {M.~S.}\ \bibnamefont {Shur}},\ eds.,\
  \href@noop {} {\emph {\bibinfo {title} {Properties of Advanced Semiconductor
  Materials: GaN, AIN, InN, BN, SiC, SiGe}}},\ \bibinfo {edition} {1st}\ ed.\
  (\bibinfo  {publisher} {Wiley-Interscience},\ \bibinfo {address} {New York},\
  \bibinfo {year} {2001})\BibitemShut {NoStop}%
\bibitem [{\citenamefont {Monkhorst}\ and\ \citenamefont
  {Pack}(1976)}]{PhysRevB.13.5188}%
  \BibitemOpen
  \bibfield  {author} {\bibinfo {author} {\bibfnamefont {H.~J.}\ \bibnamefont
  {Monkhorst}}\ and\ \bibinfo {author} {\bibfnamefont {J.~D.}\ \bibnamefont
  {Pack}},\ }\href {\doibase 10.1103/PhysRevB.13.5188} {\bibfield  {journal}
  {\bibinfo  {journal} {Phys. Rev. B}\ }\textbf {\bibinfo {volume} {13}},\
  \bibinfo {pages} {5188} (\bibinfo {year} {1976})}\BibitemShut {NoStop}%
\bibitem [{\citenamefont {Methfessel}\ and\ \citenamefont
  {Paxton}(1989)}]{PhysRevB.40.3616}%
  \BibitemOpen
  \bibfield  {author} {\bibinfo {author} {\bibfnamefont {M.}~\bibnamefont
  {Methfessel}}\ and\ \bibinfo {author} {\bibfnamefont {A.~T.}\ \bibnamefont
  {Paxton}},\ }\href {\doibase 10.1103/PhysRevB.40.3616} {\bibfield  {journal}
  {\bibinfo  {journal} {Phys. Rev. B}\ }\textbf {\bibinfo {volume} {40}},\
  \bibinfo {pages} {3616} (\bibinfo {year} {1989})}\BibitemShut {NoStop}%
\bibitem [{\citenamefont {Perdew}\ \emph {et~al.}(1996)\citenamefont {Perdew},
  \citenamefont {Burke},\ and\ \citenamefont
  {Ernzerhof}}]{PhysRevLett.77.3865}%
  \BibitemOpen
  \bibfield  {author} {\bibinfo {author} {\bibfnamefont {J.~P.}\ \bibnamefont
  {Perdew}}, \bibinfo {author} {\bibfnamefont {K.}~\bibnamefont {Burke}}, \
  and\ \bibinfo {author} {\bibfnamefont {M.}~\bibnamefont {Ernzerhof}},\ }\href
  {\doibase 10.1103/PhysRevLett.77.3865} {\bibfield  {journal} {\bibinfo
  {journal} {Phys. Rev. Lett.}\ }\textbf {\bibinfo {volume} {77}},\ \bibinfo
  {pages} {3865} (\bibinfo {year} {1996})}\BibitemShut {NoStop}%
\bibitem [{\citenamefont {Giannozzi}\ \emph {et~al.}(2009)\citenamefont
  {Giannozzi}, \citenamefont {Baroni}, \citenamefont {Bonini}, \citenamefont
  {Calandra}, \citenamefont {Car}, \citenamefont {Cavazzoni}, \citenamefont
  {Ceresoli}, \citenamefont {Chiarotti}, \citenamefont {Cococcioni},
  \citenamefont {Dabo}, \citenamefont {{Dal Corso}}, \citenamefont
  {de~Gironcoli}, \citenamefont {Fabris}, \citenamefont {Fratesi},
  \citenamefont {Gebauer}, \citenamefont {Gerstmann}, \citenamefont
  {Gougoussis}, \citenamefont {Kokalj}, \citenamefont {Lazzeri}, \citenamefont
  {Martin-Samos}, \citenamefont {Marzari}, \citenamefont {Mauri}, \citenamefont
  {Mazzarello}, \citenamefont {Paolini}, \citenamefont {Pasquarello},
  \citenamefont {Paulatto}, \citenamefont {Sbraccia}, \citenamefont {Scandolo},
  \citenamefont {Sclauzero}, \citenamefont {Seitsonen}, \citenamefont
  {Smogunov}, \citenamefont {Umari},\ and\ \citenamefont
  {Wentzcovitch}}]{QE-2009}%
  \BibitemOpen
  \bibfield  {author} {\bibinfo {author} {\bibfnamefont {P.}~\bibnamefont
  {Giannozzi}}, \bibinfo {author} {\bibfnamefont {S.}~\bibnamefont {Baroni}},
  \bibinfo {author} {\bibfnamefont {N.}~\bibnamefont {Bonini}}, \bibinfo
  {author} {\bibfnamefont {M.}~\bibnamefont {Calandra}}, \bibinfo {author}
  {\bibfnamefont {R.}~\bibnamefont {Car}}, \bibinfo {author} {\bibfnamefont
  {C.}~\bibnamefont {Cavazzoni}}, \bibinfo {author} {\bibfnamefont
  {D.}~\bibnamefont {Ceresoli}}, \bibinfo {author} {\bibfnamefont {G.~L.}\
  \bibnamefont {Chiarotti}}, \bibinfo {author} {\bibfnamefont {M.}~\bibnamefont
  {Cococcioni}}, \bibinfo {author} {\bibfnamefont {I.}~\bibnamefont {Dabo}},
  \bibinfo {author} {\bibfnamefont {A.}~\bibnamefont {{Dal Corso}}}, \bibinfo
  {author} {\bibfnamefont {S.}~\bibnamefont {de~Gironcoli}}, \bibinfo {author}
  {\bibfnamefont {S.}~\bibnamefont {Fabris}}, \bibinfo {author} {\bibfnamefont
  {G.}~\bibnamefont {Fratesi}}, \bibinfo {author} {\bibfnamefont
  {R.}~\bibnamefont {Gebauer}}, \bibinfo {author} {\bibfnamefont
  {U.}~\bibnamefont {Gerstmann}}, \bibinfo {author} {\bibfnamefont
  {C.}~\bibnamefont {Gougoussis}}, \bibinfo {author} {\bibfnamefont
  {A.}~\bibnamefont {Kokalj}}, \bibinfo {author} {\bibfnamefont
  {M.}~\bibnamefont {Lazzeri}}, \bibinfo {author} {\bibfnamefont
  {L.}~\bibnamefont {Martin-Samos}}, \bibinfo {author} {\bibfnamefont
  {N.}~\bibnamefont {Marzari}}, \bibinfo {author} {\bibfnamefont
  {F.}~\bibnamefont {Mauri}}, \bibinfo {author} {\bibfnamefont
  {R.}~\bibnamefont {Mazzarello}}, \bibinfo {author} {\bibfnamefont
  {S.}~\bibnamefont {Paolini}}, \bibinfo {author} {\bibfnamefont
  {A.}~\bibnamefont {Pasquarello}}, \bibinfo {author} {\bibfnamefont
  {L.}~\bibnamefont {Paulatto}}, \bibinfo {author} {\bibfnamefont
  {C.}~\bibnamefont {Sbraccia}}, \bibinfo {author} {\bibfnamefont
  {S.}~\bibnamefont {Scandolo}}, \bibinfo {author} {\bibfnamefont
  {G.}~\bibnamefont {Sclauzero}}, \bibinfo {author} {\bibfnamefont {A.~P.}\
  \bibnamefont {Seitsonen}}, \bibinfo {author} {\bibfnamefont {A.}~\bibnamefont
  {Smogunov}}, \bibinfo {author} {\bibfnamefont {P.}~\bibnamefont {Umari}}, \
  and\ \bibinfo {author} {\bibfnamefont {R.~M.}\ \bibnamefont {Wentzcovitch}},\
  }\href {http://www.quantum-espresso.org} {\bibfield  {journal} {\bibinfo
  {journal} {J. Phys.: Condens. Matter}\ }\textbf {\bibinfo {volume} {21}},\
  \bibinfo {pages} {395502} (\bibinfo {year} {2009})}\BibitemShut {NoStop}%
\bibitem [{\citenamefont {Bader}(1990)}]{Bader}%
  \BibitemOpen
  \bibfield  {author} {\bibinfo {author} {\bibfnamefont {R.~F.~W.}\
  \bibnamefont {Bader}},\ }\href@noop {} {\emph {\bibinfo {title} {Atoms in
  molecules: a quantum theory}}}\ (\bibinfo  {publisher} {Oxford University
  Press},\ \bibinfo {address} {New York},\ \bibinfo {year} {1990})\BibitemShut
  {NoStop}%
\bibitem [{\citenamefont {Giovannetti}\ \emph {et~al.}(2007)\citenamefont
  {Giovannetti}, \citenamefont {Khomyakov}, \citenamefont {Brocks},
  \citenamefont {Kelly},\ and\ \citenamefont {van~den
  Brink}}]{PhysRevB.76.073103}%
  \BibitemOpen
  \bibfield  {author} {\bibinfo {author} {\bibfnamefont {G.}~\bibnamefont
  {Giovannetti}}, \bibinfo {author} {\bibfnamefont {P.~A.}\ \bibnamefont
  {Khomyakov}}, \bibinfo {author} {\bibfnamefont {G.}~\bibnamefont {Brocks}},
  \bibinfo {author} {\bibfnamefont {P.~J.}\ \bibnamefont {Kelly}}, \ and\
  \bibinfo {author} {\bibfnamefont {J.}~\bibnamefont {van~den Brink}},\ }\href
  {\doibase 10.1103/PhysRevB.76.073103} {\bibfield  {journal} {\bibinfo
  {journal} {Phys. Rev. B}\ }\textbf {\bibinfo {volume} {76}},\ \bibinfo
  {pages} {073103} (\bibinfo {year} {2007})}\BibitemShut {NoStop}%
\bibitem [{\citenamefont {Ponomarenko}\ \emph {et~al.}(2013)\citenamefont
  {Ponomarenko}, \citenamefont {Belle}, \citenamefont {Jalil}, \citenamefont
  {Britnell}, \citenamefont {Gorbachev}, \citenamefont {Geim}, \citenamefont
  {Novoselov}, \citenamefont {Castro~Neto}, \citenamefont {Eaves},\ and\
  \citenamefont {Katsnelson}}]{JAP.113.136502}%
  \BibitemOpen
  \bibfield  {author} {\bibinfo {author} {\bibfnamefont {L.~A.}\ \bibnamefont
  {Ponomarenko}}, \bibinfo {author} {\bibfnamefont {B.~D.}\ \bibnamefont
  {Belle}}, \bibinfo {author} {\bibfnamefont {R.}~\bibnamefont {Jalil}},
  \bibinfo {author} {\bibfnamefont {L.}~\bibnamefont {Britnell}}, \bibinfo
  {author} {\bibfnamefont {R.~V.}\ \bibnamefont {Gorbachev}}, \bibinfo {author}
  {\bibfnamefont {A.~K.}\ \bibnamefont {Geim}}, \bibinfo {author}
  {\bibfnamefont {K.~S.}\ \bibnamefont {Novoselov}}, \bibinfo {author}
  {\bibfnamefont {A.~H.}\ \bibnamefont {Castro~Neto}}, \bibinfo {author}
  {\bibfnamefont {L.}~\bibnamefont {Eaves}}, \ and\ \bibinfo {author}
  {\bibfnamefont {M.~I.}\ \bibnamefont {Katsnelson}},\ }\href {\doibase
  http://dx.doi.org/10.1063/1.4795542} {\bibfield  {journal} {\bibinfo
  {journal} {J. Appl. Phys.}\ }\textbf {\bibinfo {volume} {113}},\ \bibinfo
  {pages} {136502} (\bibinfo {year} {2013})}\BibitemShut {NoStop}%
\bibitem [{\citenamefont {Kumar}\ \emph {et~al.}(2012)\citenamefont {Kumar},
  \citenamefont {Seol},\ and\ \citenamefont {Guo}}]{APL.101.033503}%
  \BibitemOpen
  \bibfield  {author} {\bibinfo {author} {\bibfnamefont {S.~B.}\ \bibnamefont
  {Kumar}}, \bibinfo {author} {\bibfnamefont {G.}~\bibnamefont {Seol}}, \ and\
  \bibinfo {author} {\bibfnamefont {J.}~\bibnamefont {Guo}},\ }\href {\doibase
  10.1063/1.4737394} {\bibfield  {journal} {\bibinfo  {journal} {Appl. Phys.
  Lett.}\ }\textbf {\bibinfo {volume} {101}},\ \bibinfo {eid} {033503}
  (\bibinfo {year} {2012})}\BibitemShut {NoStop}%
\bibitem [{\citenamefont {Fiori}\ \emph {et~al.}(2013)\citenamefont {Fiori},
  \citenamefont {Bruzzone},\ and\ \citenamefont {Iannaccone}}]{IEEE.60.268}%
  \BibitemOpen
  \bibfield  {author} {\bibinfo {author} {\bibfnamefont {G.}~\bibnamefont
  {Fiori}}, \bibinfo {author} {\bibfnamefont {S.}~\bibnamefont {Bruzzone}}, \
  and\ \bibinfo {author} {\bibfnamefont {G.}~\bibnamefont {Iannaccone}},\
  }\href {\doibase 10.1109/TED.2012.2226464} {\bibfield  {journal} {\bibinfo
  {journal} {IEEE Trans. Electron. Dev.}\ }\textbf {\bibinfo {volume} {60}},\
  \bibinfo {pages} {268} (\bibinfo {year} {2013})}\BibitemShut {NoStop}%
\bibitem [{\citenamefont {King-Smith}\ and\ \citenamefont
  {Vanderbilt}(1993)}]{PhysRevB.47.1651}%
  \BibitemOpen
  \bibfield  {author} {\bibinfo {author} {\bibfnamefont {R.~D.}\ \bibnamefont
  {King-Smith}}\ and\ \bibinfo {author} {\bibfnamefont {D.}~\bibnamefont
  {Vanderbilt}},\ }\href {\doibase 10.1103/PhysRevB.47.1651} {\bibfield
  {journal} {\bibinfo  {journal} {Phys. Rev. B}\ }\textbf {\bibinfo {volume}
  {47}},\ \bibinfo {pages} {1651} (\bibinfo {year} {1993})}\BibitemShut
  {NoStop}%
\bibitem [{\citenamefont {Resta}(1994)}]{RevModPhys.66.899}%
  \BibitemOpen
  \bibfield  {author} {\bibinfo {author} {\bibfnamefont {R.}~\bibnamefont
  {Resta}},\ }\href {\doibase 10.1103/RevModPhys.66.899} {\bibfield  {journal}
  {\bibinfo  {journal} {Rev. Mod. Phys.}\ }\textbf {\bibinfo {volume} {66}},\
  \bibinfo {pages} {899} (\bibinfo {year} {1994})}\BibitemShut {NoStop}%
\bibitem [{\citenamefont {Stengel}\ and\ \citenamefont
  {Spaldin}(2007)}]{PhysRevB.75.205121}%
  \BibitemOpen
  \bibfield  {author} {\bibinfo {author} {\bibfnamefont {M.}~\bibnamefont
  {Stengel}}\ and\ \bibinfo {author} {\bibfnamefont {N.~A.}\ \bibnamefont
  {Spaldin}},\ }\href {\doibase 10.1103/PhysRevB.75.205121} {\bibfield
  {journal} {\bibinfo  {journal} {Phys. Rev. B}\ }\textbf {\bibinfo {volume}
  {75}},\ \bibinfo {pages} {205121} (\bibinfo {year} {2007})}\BibitemShut
  {NoStop}%
\bibitem [{\citenamefont {Stengel}\ \emph {et~al.}(2009)\citenamefont
  {Stengel}, \citenamefont {Vanderbilt},\ and\ \citenamefont
  {Spaldin}}]{PhysRevB.80.224110}%
  \BibitemOpen
  \bibfield  {author} {\bibinfo {author} {\bibfnamefont {M.}~\bibnamefont
  {Stengel}}, \bibinfo {author} {\bibfnamefont {D.}~\bibnamefont {Vanderbilt}},
  \ and\ \bibinfo {author} {\bibfnamefont {N.~A.}\ \bibnamefont {Spaldin}},\
  }\href {\doibase 10.1103/PhysRevB.80.224110} {\bibfield  {journal} {\bibinfo
  {journal} {Phys. Rev. B}\ }\textbf {\bibinfo {volume} {80}},\ \bibinfo
  {pages} {224110} (\bibinfo {year} {2009})}\BibitemShut {NoStop}%
\bibitem [{\citenamefont {Giustino}\ and\ \citenamefont
  {Pasquarello}(2005)}]{PhysRevB.71.144104}%
  \BibitemOpen
  \bibfield  {author} {\bibinfo {author} {\bibfnamefont {F.}~\bibnamefont
  {Giustino}}\ and\ \bibinfo {author} {\bibfnamefont {A.}~\bibnamefont
  {Pasquarello}},\ }\href@noop {} {\bibfield  {journal} {\bibinfo  {journal}
  {Phys. Rev. B}\ }\textbf {\bibinfo {volume} {71}},\ \bibinfo {pages} {144104}
  (\bibinfo {year} {2005})}\BibitemShut {NoStop}%
\bibitem [{\citenamefont {Wu}\ \emph {et~al.}(2006)\citenamefont {Wu},
  \citenamefont {Di\'eguez}, \citenamefont {Rabe},\ and\ \citenamefont
  {Vanderbilt}}]{PhysRevLett.97.107602}%
  \BibitemOpen
  \bibfield  {author} {\bibinfo {author} {\bibfnamefont {X.}~\bibnamefont
  {Wu}}, \bibinfo {author} {\bibfnamefont {O.}~\bibnamefont {Di\'eguez}},
  \bibinfo {author} {\bibfnamefont {K.~M.}\ \bibnamefont {Rabe}}, \ and\
  \bibinfo {author} {\bibfnamefont {D.}~\bibnamefont {Vanderbilt}},\ }\href
  {\doibase 10.1103/PhysRevLett.97.107602} {\bibfield  {journal} {\bibinfo
  {journal} {Phys. Rev. Lett.}\ }\textbf {\bibinfo {volume} {97}},\ \bibinfo
  {pages} {107602} (\bibinfo {year} {2006})}\BibitemShut {NoStop}%
\bibitem [{\citenamefont {Marzari}\ and\ \citenamefont
  {Vanderbilt}(1997)}]{PhysRevB.56.12847}%
  \BibitemOpen
  \bibfield  {author} {\bibinfo {author} {\bibfnamefont {N.}~\bibnamefont
  {Marzari}}\ and\ \bibinfo {author} {\bibfnamefont {D.}~\bibnamefont
  {Vanderbilt}},\ }\href {\doibase 10.1103/PhysRevB.56.12847} {\bibfield
  {journal} {\bibinfo  {journal} {Phys. Rev. B}\ }\textbf {\bibinfo {volume}
  {56}},\ \bibinfo {pages} {12847} (\bibinfo {year} {1997})}\BibitemShut
  {NoStop}%
\bibitem [{\citenamefont {Sgiarovello}\ \emph {et~al.}(2001)\citenamefont
  {Sgiarovello}, \citenamefont {Peressi},\ and\ \citenamefont
  {Resta}}]{PhysRevB.64.115202}%
  \BibitemOpen
  \bibfield  {author} {\bibinfo {author} {\bibfnamefont {C.}~\bibnamefont
  {Sgiarovello}}, \bibinfo {author} {\bibfnamefont {M.}~\bibnamefont
  {Peressi}}, \ and\ \bibinfo {author} {\bibfnamefont {R.}~\bibnamefont
  {Resta}},\ }\href {\doibase 10.1103/PhysRevB.64.115202} {\bibfield  {journal}
  {\bibinfo  {journal} {Phys. Rev. B}\ }\textbf {\bibinfo {volume} {64}},\
  \bibinfo {pages} {115202} (\bibinfo {year} {2001})}\BibitemShut {NoStop}%
\bibitem [{\citenamefont {Nakamura}\ \emph {et~al.}(2006)\citenamefont
  {Nakamura}, \citenamefont {Ishihara},\ and\ \citenamefont
  {Natori}}]{JAP.99.054309}%
  \BibitemOpen
  \bibfield  {author} {\bibinfo {author} {\bibfnamefont {J.}~\bibnamefont
  {Nakamura}}, \bibinfo {author} {\bibfnamefont {S.}~\bibnamefont {Ishihara}},
  \ and\ \bibinfo {author} {\bibfnamefont {A.}~\bibnamefont {Natori}},\
  }\href@noop {} {\bibfield  {journal} {\bibinfo  {journal} {J. Appl. Phys.}\
  }\textbf {\bibinfo {volume} {99}},\ \bibinfo {pages} {054309} (\bibinfo
  {year} {2006})}\BibitemShut {NoStop}%
\end{thebibliography}
%
%%%%%%%%%%%%%%%%%%%%%%%%%%%%%%%%%%%%%%

%merlin.mbs apsrev4-1.bst 2010-07-25 4.21a (PWD, AO, DPC) hacked
%Control: key (0)
%Control: author (72) initials jnrlst
%Control: editor formatted (1) identically to author
%Control: production of article title (-1) disabled
%Control: page (0) single
%Control: year (1) truncated
%Control: production of eprint (0) enabled
%

\end{document}